\documentclass[12pt]{article}
\usepackage{times}
\usepackage{epsfig,graphicx}
\usepackage{amsmath}

\begin{document}
\title{Epidemic spreading with time delay \\ in complex networks}
\author{Xin-Jian Xu$^{1}$, Hai-Ou Peng$^{1}$, Xiao-Mei Wang$^{2}$, and Ying-Hai Wang$^{1}$\\
$^{1}$Institute of Theoretical Physics, Lanzhou University\\ Lanzhou Gansu 730000, China\\
$^{2}$Herbert Irving Comprehensive Cancer Center, Columbia
University\\ New York 10032, USA}

\maketitle

\begin{abstract}
We present a modified \emph{susceptible-infected-susceptible}
(SIS) model on complex networks, small-world and scale-free, to
study epidemic spreading with the effect of time delay which is
introduced to the infected phase. Considering topologies of the
networks, both uniform and degree-dependent delays are studied
during the contagion process. It is found that the existence of
delay will enhance both outbreaks and prevalence of infectious
diseases in the networks.
\end{abstract}

\section{Introduction}

Complex weblike structures describe a wide variety of systems of
high technological and intellectual importance and have attracted
an increasing interest recently \cite {Albert_1, Dorogovtsev_1,
Newman_1}. The explosion of the general interest in the problem of
the structure and evolution of most different networks is mainly
connected with two characters, the small average path lengths
among any two nodes (small-world property) \cite {Watts_1} and a
power law distribution (scale-free property), $P(k)\sim k^{-
\gamma}$ with $2 \leq \gamma \leq 3$, for the probability that any
node has $k$ connections to other nodes \cite {Barabasi_1}.

In the study of complex networks, a good example is to inspect the
effect of their complex features on the dynamics of epidemic and
disease spreading. It is easy to foresee that the characterization
and understanding of epidemic dynamics on these networks can find
immediate applications to a large number of problems, such as
computer virus infections \cite {Kephart}, epidemiology \cite
{Bailey}, and the spreading of polluting agents \cite {Hill}, etc.
Recent papers \cite {May, Kuperman, Pastor, Newman_2, Moreno,
Boguna, Boguna_1} have given some valuable insights of that: for
small-world networks, there is a critical threshold below which an
infection with a spreading rate dies out; on the contrary, for
scale-free networks, even an infection with a low spreading rate
will prevalence the entire population.

In many social and biological systems, however, temporal delay is
natural and the finite time interval required for the information
transmission between two elements may be important \cite {Traub,
Gray, Foss, Li}. In this paper we will introduce time delay to the
standard SIS model \cite {Bailey} on two prototype complex
networks, the Watts-Strogatz (WS) model and the
Barab\'{a}si-Albert (BA) model. Which is motivated by the
following questions: during the process of epidemic spreading, if
an individual is infected there is always a period of time before
he (or she) becomes recovery, including the time an infected
individual is found and sent to a hospital, and the time a patient
is being cured, etc.

The paper is organized as follows. In Sec. II we first define the
model with time delay on complex networks. Then we discuss the
uniform delay in Sec. III and the degree-dependent delay in Sec.
IV. Finally we draw our conclusions and perspectives in Sec. V.

\section{The Model}

In this section, we shall introduce the effect of time delay to
the standard SIS model on complex networks, in which each node
represents an individual of the population and the edges represent
the physical interactions through which an infection spreads. The
two prototype complex networks, WS graph and BA graph, can be
constructed as follows.

\emph{WS graph}: Starting with a ring of $N$ vertices, each
connected to its $2K$ nearest neighbors by undirected edges, and
then each local link is visited once with the rewiring probability
$p$ it is removed and reconnected to a randomly chosen node.
Duplicate and self-connected edges are forbidden. After the whole
sweep of the entire network, a small-world graph is constructed
with an average connectivity $\langle k \rangle = 2K$ (in the
present work we will consider the parameters $N=10^{5}$, $p=0.1$
and $K=5$).

\emph{BA graph}: Starting from a small number $m_{0}$ of nodes,
every time step a new vertex is added, with $m$ links that are
connected to an old node $i$ with probability
$\Pi_{i}=k_{i}/\sum_{j}k_{j}$, where $k_{i}$ is the connectivity
of the $i$th node. After iterating this scheme a sufficient number
of times, we obtain a network composed by $N$ nodes with
connectivity distribution $P(k) \sim k^{-3}$ and average
connectivity $\langle k \rangle =2m$  (in the present work we will
consider the parameters $N=10^{5}$, $m_{0}=10$ and $m=5$).

In our model, an individual is described by a single dynamical
variable adopting one of the two stages: \emph{susceptible} and
\emph{infected}. The two states completely neglect the details of
the infection mechanism within each individual. The transmission
of the disease is described in an effective way with the following
rules: A susceptible individual at time $t$ will pass to the
infected state with the rate $\nu$ at time $t+ \Delta t$ if it is
connected to one or more infected individuals, where $\Delta t$ is
the time step of Monte Carlo (MC) simulations. Infected
individuals at time $t$ will pass to the susceptible state again
with the rate $\delta$ at time $t+ \Delta t + \tau_{I}$, where
$\tau_{I}$ denotes the delay time in the infected phase. Here, an
effective spreading rate $\lambda = \nu / \delta$ is defined. We
can still keep the generality by setting $\delta = 1$. Individuals
run stochastically through the cycle, \emph{susceptible}
$\rightarrow$ \emph{infected} $\rightarrow$ \emph{susceptible}.

In the present work, we have performed MC simulations of the model
with synchronously updating in the network. Initially, the number
of infected nodes is $5$ percent of the size of the network. The
total sampling times are $10000$ MC time steps. After appropriate
relaxation times, the systems stabilize in a steady state.
Simulations were implemented on the networks averaging over $100$
different realizations. Given a network, an important observable
is the prevalence $\rho$, which is the time average of the
fraction of infected individuals reached after a transient from
the initial condition (averaging over $1000$ time steps in this
context). The information on the global spreading of infected
diseases is contained in the function $\rho(\lambda, \tau_{I})$.

\section{Uniform delay}

We firstly consider that all the individuals in the network have
an uniform delay
\begin{equation}
\tau_I^{i} = \tau, \tag{1}\label{eq1}
\end{equation}
that is to say, the details of the delay mechanism within
individuals are independent of the connectivity fluctuations of
the networks. In our simulations, the system stabilizes in dynamic
equilibrium after appropriate relaxation times.

\begin{figure}
\centerline{\epsfxsize=12cm \epsffile{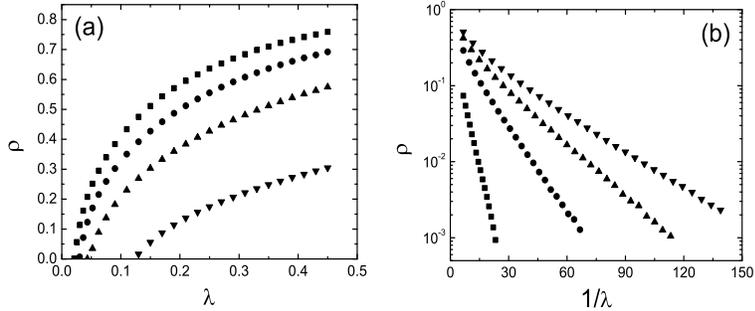}}

\caption{Plots of $\rho$ vs $\lambda$ in the WS (a) and $\rho$ vs
$1/\lambda$ in the BA (b) networks under different values of the
uniform delay time (from bottom to top) $\tau = 0$, $2$, $4$, and
$6$, respectively.} \label{fig1}
\end{figure}

In Fig. \ref{fig1}, under the different values of delay time, the
plots of $\rho$ versus $\lambda$ in the WS and $\rho$ versus
$1/\lambda$ in the BA networks are shown. In the case of $\tau =
0$, the model becomes the standard SIS model and  gives an
epidemic threshold $\lambda_{c} \sim 1/\langle k \rangle$ in the
WS network and $\rho \sim \exp(-1/m\lambda)$ in the BA network,
which were firstly introduced by Pastor-Satorras and Vespignani
\cite {Pastor} by using mean-field theory. In the presence of time
delay ($\tau=2$, $4$, $6$), one can easily find that the epidemic
prevalence in steady state increases greatly, which induce the
epidemic threshold $\lambda_{c}$ to becomes smaller in the WS
network and the scaling effect to become weaker in the BA network.

\begin{figure}
\centerline{ \epsfxsize=12cm \epsffile{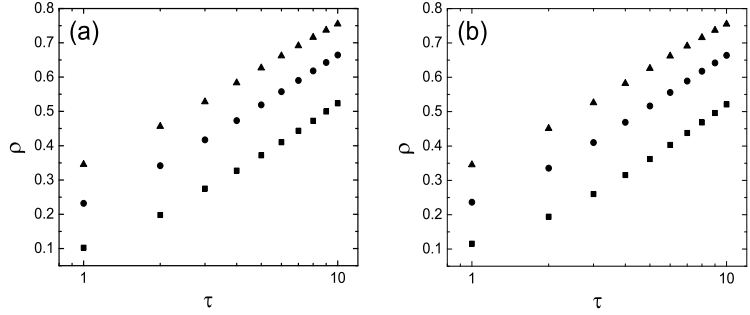}}

\caption{Linear-log plots of densities of infected nodes $\rho$ vs
$\tau$ in the WS (a) and BA (b) networks under different values of
the effective spreading rate (from bottom to top) $\lambda =
0.10$, $0.15$, and $0.20$, respectively.} \label{fig2}
\end{figure}

In Fig. \ref{fig2} we show the linear-log plots of $\rho$ versus
$\tau$ in the WS and BA networks under different values of the
effective spreading rate, $\lambda = 0.10$ (full squares), $0.15$
(full circles), and $0.20$ (full triangles), respectively.
Consistently, the epidemic prevalence in steady state increases
with the enhancement of the effect of time delay. In addition,
both WS and BA networks present a linear relation between $\rho$
and $\ln \tau$, that is, $\rho \sim A +B\ln \tau $, for given
effective spreading rate. Simulations indicate that the parameter
$B$ is identical for both networks.

\begin{figure}
\centerline{ \epsfxsize=8cm \epsffile{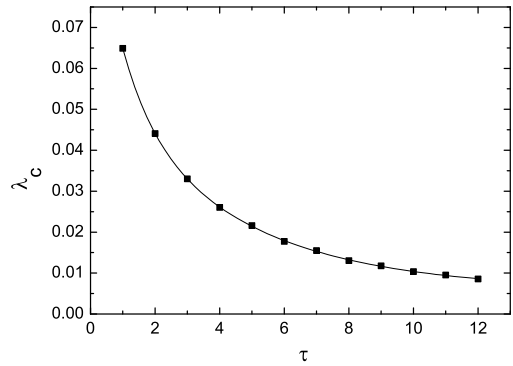}}

\caption{The plot of the epidemic threshold $\lambda_{c}$ as a
function of uniform delay time $\tau$ in the WS network. The solid
line is a fit to the form $\lambda_{c}=C+De^{-\tau/\tau_{0}}$.}
\label{fig3}
\end{figure}

To find the relation between the epidemic threshold and time delay
in the WS network, we plot $\lambda_{c}$ as a function of $\tau$
in Fig. \ref{fig3}. Closed squares represent the numerical results
and the solid line is a fit to the form $\lambda_{c}(\tau) = C + D
e^{-\tau/\tau_{0}}$, which implies there is a relation of the
first order exponential decay between $\lambda_c$ and $\tau$.
Parameters values (given by simulations) $C=0.0113 \pm 0.0007$,
$D=0.0713 \pm 0.0002$, and $\tau_{0}=2.14 \pm 0.02$.

\section{Degree-dependent delay}

In all the simulations above, we take homogeneous individual
activities in the networks, i.e., the delay is identical for each
individual during the evolution of the system. However,
considering the heterogeneousness of networks, the distribution of
the connectivity, we suggest a degree-dependent delay form
\begin{equation}
\tau_{I}^{i} = \frac{k_{i}^{-\alpha}}{\alpha}, \tag{2}\label{eq2}
\end{equation}
where $\alpha$ is a tunable parameter. In Eq. (\ref{eq2}), the
delay time is inversely proportional to $k_{i}$, namely, the
larger degree a node has, the smaller the delay time the node
takes. In language of sociology, $k_{i}$ represents the degree of
the activity of an individual. So active individuals are easier to
be found if they are infected and it will take less time for them
to become susceptible again. Here, the details of the recovery
mechanism within each individual are completely neglected.

\begin{figure}
\centerline{\epsfxsize=12cm \epsffile{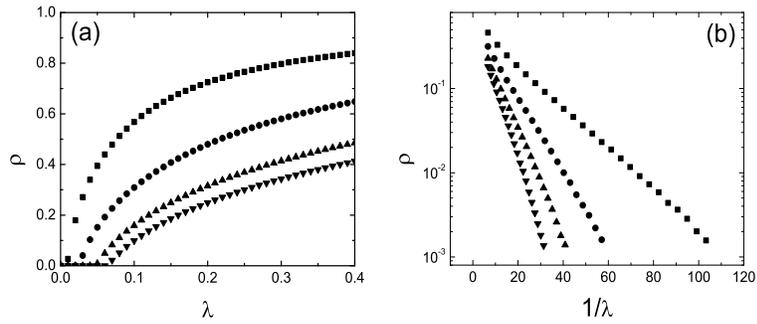}}

\caption{Plots of $\rho$ vs $\lambda$ in the WS (a) and $\rho$ vs
$1/\lambda$ in the BA (b) networks under different values of the
degree-dependent delay time. Parameter values (from bottom to top)
$\alpha = 0.15$, $0.25$, $0.35$, and $0.45$, respectively.}
\label{fig4}
\end{figure}

\begin{figure}
\centerline{\epsfxsize=8cm \epsffile{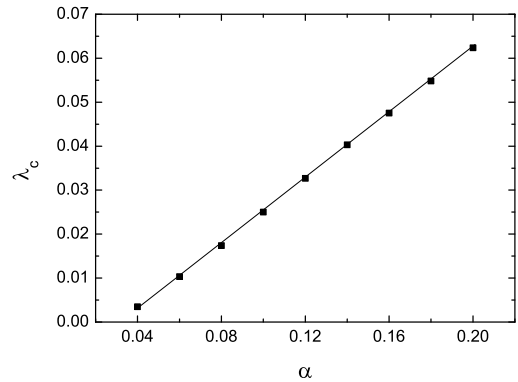}}

\caption{The plot of the epidemic threshold $\lambda_c$ as a
function of the tunable parameter $\alpha$ in the WS network. The
solid line is a fit to the form $\lambda_{c}(\alpha) \sim E + F
\alpha$.} \label{fig5}
\end{figure}

We perform simulations of the model with the same rule and the
system reaches a dynamic equilibrium after an initial transient.
In Fig. \ref{fig4}, under the different values of $\alpha$, the
plots of $\rho$ versus $\lambda$ in the WS and $\rho$ versus
$1/\lambda$ in the BA networks are shown. With the reducing of the
tunable parameter (from $0.45$ to $0.15$), the epidemic threshold
$\lambda_{c}$ becomes smaller in the WS network and the scaling
effect become weaker in the BA network. The results are
qualitatively consistent with the case of uniform delay since the
value of delay time is inverse proportional to the tunable
parameter $\alpha$ (see Eq. (\ref{eq2})).

At the end, the plot of epidemic threshold $\lambda_{c}$ as a
function of $\alpha$ in the WS network is shown in Fig.
\ref{fig5}. Closed squares represent the numerical results and the
solid line is a fit to the form $\lambda_{c}(\alpha) = E + F
\alpha$, which predicts a linear relation between $\lambda_{c}$
and $\alpha$. Parameters values (given by simulations) $E=-0.010
\pm 0.002$, $F=0.37 \pm 0.01$.

\section{Conclusions}

We have investigated the spread of infectious diseases with time
delay in complex networks. The effect is presented in the infected
phase of the standard SIS model. Both the uniform and
degree-dependent delays are considered during the contagion
process. It was found that the existence of delay will enhance
both outbreaks and prevalence of infectious diseases in the
networks. However, the results are based on numerical simulations.
It deserves to make further study on the theoretical side and
explore the connection to real data of diseases. In reality, there
always exists a mean incubation period during the spread of
epidemics \cite {Riley, Small}. Our model may provide an
explanation for this spreading phenomenon in social systems.

Since time delay arises naturally from the kinetic theory \cite
{Jou, Chen}, physicists can contribute to topics related to that,
such as the explanation of mutant virus strains \cite {Lee}, the
modelling of the front shapes in virus infections \cite {Yin}, and
the characterization of the speed of virus infections \cite
{Fort}, etc.

\section*{Acknowledgements}
We thank Yong Chen for valuable discussions. This work was
supported by the Doctoral Research Foundation awarded by Lanzhou
University.

\newpage

\bigskip

\end{document}